\documentstyle[preprint,aps]{revtex}
\begin{document}
\draft
\title{An improved calculation of the third virial coefficient\\
of a free anyon gas}
\author{J. Law\cite{JL}, Avinash Khare\cite{AK}, R. K. Bhaduri
and Akira Suzuki\cite{AS}}
\address{Department of Physics and Astronomy, Mcmaster University,\\
 Hamilton, Ontario, Canada L8S 4M1.}

\date{\today}
\maketitle
\begin{abstract}
For three anyons confined in a harmonic oscillator, only
the class of states that interpolate nonlinearly with the
statistical parameter contributes to the third virial
coefficient of a free anyon gas. Rather than evaluating the full
three-body partition function as was done in an earlier
publication (Phys. Rev.\ {\bf A46}, 4693 (1992) ), here only the
nonlinear contribution is calculated, thus avoiding delicate
cancellations between the irrelevant linear part and the
two-body partition function. Our numerical results are
consistent with the simple analytical form suggested recently by
Myrheim and Olaussen.
\end{abstract}


\narrowtext

In this note we report an improved numerical calculation of the third
virial coefficient of a free anyon gas using the method of the
hyperspherical coordinates of an earlier publication\cite{LSB92}. The virial
expansion of the two-dimensional anyon gas may be written as
\begin{equation}
{{P\beta}\over{\rho}}=1+A_2(\alpha)(\rho\lambda^2)%
              +A_3(\alpha)(\rho\lambda^2)^2+...,
\end{equation}
where $P,\rho$ and $\beta$ are the pressure density and the inverse
temperature in units of the Boltzmann constant,
and $\lambda=\hbar (2\pi\beta/m)^{1/2}$ is the thermal wavelength. The
expansion parameter is the dimensionless quantity ($\rho\lambda^2$), and
$A_2, A_3$, are the second and third virial coefficients of the gas, as
a function of the dimensionless statistical parameter $\alpha$. Note
that the coefficients $A_2, A_3$ etc. are temperature independent, since
there is no length scale in the statistical interaction. To evaluate the
virial coefficients it is convenient to confine the anyons in a
two-dimensional harmonic oscillator of frequency $\omega$, and take the
limit of the appropriate combinations of the partition functions as
$(\hbar\omega\beta) \rightarrow 0$. Denote the N-body partition function
in the harmonic oscillator by $Z_N$. The centre-of-mass part $Z_1$ may
be separated out, and the relative part is given by
$\tilde {Z}_N = Z_N/Z_1$. Then the second and the third virial
coefficients are given by ($x=\hbar\omega\beta$)
\begin{eqnarray}
A_2(\alpha)=&&\lim_{x \rightarrow 0}A_2(\alpha,x)\nonumber\\
=&&\lim_{x\rightarrow 0}x^{-2}[1-2\tilde{Z}_2/Z_1],\\
A_3(\alpha)=&&\lim_{x\rightarrow 0}A_3(\alpha,x)\nonumber\\
=&&\lim_{x\rightarrow 0}%
2x^{-4}[1+8(\tilde{Z}_2/Z_1)^2 -5(\tilde{Z}_2/Z_1)- 3(\tilde{Z}_3/Z_1^2)].
\end{eqnarray}

The calculations are performed in the fermionic basis, i.e. for
$\alpha=0$, the particles are noninteracting fermions while for
$\alpha=1$ they behave like noninteracting bosons. In Eqs.(2) and (3),
$ Z_1$ and $\tilde {Z}_2$ are analytically known ($t = exp(-x)$):
\begin{equation}
Z_1=t/(1-t)^2,~~~~~~~~~\tilde{Z_2}=t^2(t^\alpha+t^{-\alpha})/(1-t^2)^2,
\end{equation}
while $\tilde Z_3$ in Ref.\ \cite{LSB92} was calculated from the
numerically obtained spectrum of the three-anyon system. Three particles
in two dimensions, after elimination of the centre-of-mass, have four
independent degrees of freedom. The problem may be regarded as that of
one body in 4-dimensions. Moreover, it is convenient to use the
hyperspherical coordinates\cite{KL87} of a 4-dimensional sphere with
radius $R$ and three independent angles $\theta$, $\phi$ and $\psi$.
This is because in these coordinates, the fermionic (or bosonic) basis
of states may be constructed by specifying their quantum numbers using a
simple set of rules\cite{KM91} . The three-anyon Hamiltonian may be
separated in a radial and an angular part, and the eigenvalue spectrum
obtained by diagonalizing the angular part\cite{LSB92,SVZ93} of the
Hamiltonian in the basis of the hyperspherical harmonics
$Y_{N,\nu,\lambda}(\theta,\phi,\psi)$. In addition, there is another
quantum number $n'$ associated with the radial excitations. The
eigenstates of the harmonic oscillator are completely specified by the
set of quantum numbers $(N,\nu,\lambda,n')$ with eigenenergies given by
$E_{N,n'}= (N + 2n' + 2)\hbar\omega$. The form of the Hamiltonian and
other mathematical details will not be repeated here, since these are
given in Ref.\ \cite {LSB92}.

The calculation of $A_3$ in Ref.\ \cite {LSB92} involved a delicate
cancellation of terms in the square brackets on the right-hand side of
Eq.(3). Since $A_3$ is known to be finite\cite{SVZ93,MO93,dVO92}, there
should be no terms of order $x$ and $x^3$ within the bracket, and all
terms of order $x^2$ should cancel exactly as $x^2 \rightarrow 0$, if
the numerical calculations were exact. In practice, due to the
truncation of the basis and numerical inaccuracies, these are not
satisfied. The numerical calculations in Ref.\ \cite {LSB92} were
reliable for $x^2 \ge 0.8$, and careful interpolation was necessary to
reach the limit of $x^2=0$. Part of the error resulting from this
incomplete cancellation takes place between the terms involving
$\tilde {Z}_2$ and only that portion of $\tilde Z_3$ which originates
from the linearly interpolating eigenvalues of the three-body problem.
The analytical expression of this part of $\tilde Z_3$ is
known\cite{SVZ93,S91}. We may decompose $\tilde Z_3$ into two parts:
\begin{equation}
\tilde Z_3 = \tilde Z_3^{L}+\tilde Z_3^{NL},
\end{equation}
where
\begin{equation}
\tilde Z_3^{L}={{t^5(t^{3\alpha}+t^{-3\alpha})}%
               \over{(t-1)^4(t+1)^2(1+t+t^2)^2}},
\end{equation}
is the portion from the linearly interpolating states, and
$\tilde Z_3^{NL}$ is the nonlinear part. Although $\tilde Z_3^{NL}$
diverges as $x^{-4}$ for $x \rightarrow 0$, the quantity
\begin{equation}
\Delta\tilde Z_3^{NL}(\alpha)=[\tilde Z_3^{NL}(\alpha)-\tilde Z_3^{NL}(0)],
\end{equation}
is finite. Using Eqs.\ (3)-(6), it is straight forward to show that
\begin{equation}
\Delta A_3(\alpha)=A_3(\alpha)-A_3(0)=-6\lim_{x \rightarrow 0}%
\Delta\tilde{Z}_3^{NL}(\alpha).
\end{equation}
It follows from the above equation that it is only necessary to compute
$\Delta \tilde {Z}_3^{NL}(\alpha)$ numerically. This necessitates
identifying and excluding the linearly interpolating states (after
diagonalizing the three-body Hamiltonian) from the calculation of the
partition function. These linearly interpolating states change by 3
units of $\hbar\omega$ as the statistical parameter $\alpha$ varies from
0 to 1, and may be identified by the set of rules given in the
Appendix.

The diagonalization of the angular part of the Hamiltonian (as specified
in Ref.\ \cite{LSB92}) was performed with a basis truncated at
\begin{eqnarray}
N_{max} &&= 130 \text{ for } 0 \le \vert \lambda\vert  \le 20\nonumber\\
&&= 100 \text{ for } 21 \le \vert \lambda\vert  \le 72.
\end{eqnarray}
The quantum number $\lambda$ determines the angular momentum of the
three-body system. Note that this basis for diagonalization
is considerably bigger than used in Ref.\ \cite{LSB92}. After the
diagonalization, only the nonlinearly interpolating states were
used to calculate $\Delta \tilde {Z}_3^{NL}(\alpha)$ as a function of $x^2$
for various values of $\alpha$. The nodeless ($n' = 0$) states were
included in the evaluation of $\sum_i \exp(\beta E_i)$, and the infinite
tower of nodal excitations were taken into account by the multiplicative
overall factor of $[1-exp(-2x)]^{-1}$. The numerical results changed
insignificantly even if the sum over all the eigenstates, including the
tower of nodal excitations were cut off at $90\hbar\omega$. The quantity
$[-6\Delta \tilde Z_3^{NL}(\alpha)]$ is computed as a function of $x^2$
for $\alpha = 0.05$ to $0.5$ in steps of 0.05. In principle, its value
at $x^2 = 0$ shou1d yield $\Delta A_3(\alpha)$. In practice, there is
still a divergence due to numerical inaccuracies, since the cancellation
of singularities between $\tilde {Z}_3^{NL}(\alpha)$ and
$\tilde Z_3^{NL}(0)$ is not perfect as $x^2 \rightarrow 0$, since
$\tilde Z_3^{NL}(\alpha)$ is calculated less accurately than
$\tilde Z_3^{NL}(0)$. The spurious singularities in
$\Delta \tilde {Z}_3^{NL}(\alpha)$ have to be isolated to extract
$\Delta A_3(\alpha)$. To this end, we define
$\Delta A_3(\alpha,x^2) = - 6 \Delta \tilde Z_3^{NL}(\alpha)$ for any x, and
fit\cite {NR92} this function by the following form containing
six $\alpha$-dependent parameters in the range $0.05 \le x^2 \le 3.4$:
\begin{equation}
\Delta A_3(\alpha,x^2)={{a}\over{x^{2n}}}+c\ %
             \exp[ b_2x^2+b_4x^4+b_6x^6+b_8x^8].
\end{equation}
The computed $\Delta \tilde Z_3^{NL}(\alpha)$ is sensitive to the
basis size only for $x^2 \le 0.2$.

In the above fit, the coefficient $c=\Delta A_3(\alpha)$, and is found
to be numerically stable even when higher order terms
$b_{10}x^{10} + b_{12}x^{12} + b_{14}x^{14} $ are added in the exponent
of Eq.\ (10).
The parameter $n$ in Eq.(10) is close to 2 upto $\alpha=0.20$, but
decreased monotonically for larger $\alpha$, with $n=1.56$ for
$\alpha=0.50$.
The convergence of the extracted $\Delta A_3(\alpha)$
values are also tested by performing the same calculation of
$\Delta \tilde Z_3^{NL}(\alpha)$, but including only eigenvalues upto
$N_{max}=60$. The $\Delta A_3(\alpha)$ values so obtained are displayed in
Table\ 1 for various values of $\alpha$ in the range 0.05 to 0.40. For
$\alpha > 0.40$, the extracted $\Delta A_3(\alpha)$ is not too stable, and is
too inaccurate to be listed. This is because the eigenvalues of the
Hamiltonian matrix for the larger $\alpha$ values tend to be more and
more inaccurate.

Numerically, of course, only A3 values upto $\alpha = 0.5$ would
suffice, because of the relation $A_3(\alpha)=A_3(1-\alpha)$ derived
by Sen\cite {S91} . Recently, Myrheim and Olaussen\cite{MO93} have
calculated $A_3(\alpha)$ numerically using a path integral
representation of the three body partition function. Their method of
calculation is completely different, and uses the winding number
formalism rather than the statistical interaction. From their extensive
numerical work, they suggest\cite{MO93,MO93a} that perhaps
$\Delta A_3(\alpha)$ obeys the remarkably simple relation
\begin{equation}
\Delta A_3(\alpha)={\sin^2(\pi\alpha) \over 12\pi^2}.
\end{equation}
For small $\alpha$, this is in agreement with the perturbative analytical
result of $\alpha^2/12$ of
de Veigy and Ouvry\cite{dVO92}. In Fig.\ 1, the computed values
$\Delta A_3(\alpha)$ of the present calculation are compared with the
simple form given by Eq.\ (11). Our calculated $\Delta A_3(\alpha)$
values, confirm Eq.\ (11) with good accuracy upto $\alpha = 0.25$, beyond
which our numerical method starts to lose accuracy. This independent
check of the Myrheim-Olaussen calculation is the main content of this
note. Fig.\ 1 also shows that the inaccuracies for larger $\alpha$ build
up rapidly in the method of Ref.\ \cite{LSB92}, where the nonlinear part
$\tilde {Z}_{NL}$ was not isolated.

This research was supported by grants from NSERC, Canada. NATO grant
support is also acknowledge by JL.

\section*{Appendix}
The exact solutions for the linear states with slopes of
$\pm 3\hbar\omega$ have been written down by Wu\cite{W84} in Jacobi
coordinates. Using the details given in Ref.\ \cite{KM91}, we can easily
write down these solutions in hyperspherical coordinates and identify
the quantum numbers $N,\nu,\lambda,n'$ and the degeneracy of these
states. We find that for $N = (6p-3), (6p-1),$ or $(6p + 1)$ (with
$p=1,2,3...$), there are $2p^2$ degenerate states with $n'=0$, (and
hence total energy $E = N+2$), while for $N = 6p, (6p+2)$, or $(6p+4)$ (with
$p=1,2,3,...$) there are $2p(p+1)$ degenerate states with $n'=0$. For
example, for $N = (6p-3)$, one finds that the states with
$\lambda= \pm(6p-3)$ are both $p-$fold degenerate with the
corresponding $\nu$ quantum number being
$\nu=(6p-3), (6p-9), ... 3$ (note that
$\nu, N >0;\  \nu, \vert \lambda\vert  \le N;\  \nu, N, \vert \lambda\vert $
have the same parity, and $\nu$ is always a multiple of 3). For the same
$N$, the four states with $\lambda=\pm (6p-5), \pm (6p-7) $ are all
$(p-1)-$fold degenerate ($\nu=(6p-3),(6p-9), ... 9$).
The next four states with
$\lambda=\pm (6p-9), \pm (6p-11) $ are all $(p-2)-$fold degenerate
($\nu=(6p-3),(6p-9), ... 15$), and so on. Finally the four states with
$\lambda=\pm (2p+7),\pm (2p+5)$ are all non-degenerate.
Exactly the same rules also apply when $N = (6p - 1)$ or $(6p + 1)$.
On the other hand, when $N$ is even, for  $N = 6p$, we find that
the four states with $\lambda=\pm 6p, \pm (6p-2)$, are all $p-$fold
degenerate with $\nu=6p, (6p-6), ... 3$. The next four states
with $\lambda =\pm (6p-4), \pm (6p- 6)$ are all $(p-1)-$fold degenerate and
so on. Finally the four states with $\lambda = \pm (2p+2), \pm (2p+4)$
are all non-degenerate. Similar rules also apply when $N = (6p +2)$
or $(6p + 4)$. In this way we are able to identify all the linear
states in the spectra and exclude them from the evaluation of
the partition function.

\begin{figure}
\caption{ Plot of $\Delta A_3(\alpha)$ as a function of the statistical
parameter $\alpha$ in the fermionic basis.
[a] Present calculation (top row of Table \ 1), with the six-parameter
fit of Eq.\ (10).
[b] The Variations in the extracted $\Delta A_3(\alpha)$, with a
nine-parameter fit to $\Delta A_3(\alpha,x^2)$. This is done by
including the terms $b_{10}x^{10}+b_{12}x^{12}+b_{14}x^{14}$ in the
exponent of Eq.\ (10).
[c] The plot of Eq. \ (11), (the Myrheim-Olaussen conjecture
for $\Delta A_3(\alpha)$).
[d] The calculated $\Delta A_3(\alpha)$ values by the method
given in Ref. \ [1], but in the enlarged basis, Eq.\ (9).
}
\end{figure}

\mediumtext
\begin{table}
\caption{The calculated quantity $\Delta A_3(\alpha)=[A_3(\alpha)-A_3(0)]$ is
tabulated as a function of $\alpha$. The first row of $\Delta A_3(\alpha)$
is obtained by including all the eigenenergies in the space defined by
Eq.\ (9). The six-parameter fit, as given by Eq.\ (10), is used. The
second row is computed by including only eigenvalues upto
$E_{cut}=60\hbar\omega$.}

\begin{tabular}{ccccccccc}
$\alpha$& 0.05 &0.10 & 0.15 & 0.20 & 0.25 & 0.30 & 0.35 & 0.40\\
\tableline
$\Delta A_3(\alpha)\times 10^{4}$&2.08&8.22& 17.3& 29.0&
 41.8& 53.8& 62.7& 65.7\\
\tableline
$\Delta A_3(\alpha) \times 10^{4}$&2.13&8.34& 17.6& 29.4&
 42.4& 54.7&63.9& 67.4\\
$E_{cut}=60\hbar\omega$ \\
\end{tabular}
\end{table}

\end{document}